\documentstyle[aps,epsf,psfig,twocolumn,prl]{revtex}
\pssilent

\def\be{\begin{equation}}
\def\ee{\end{equation}}
\def\ba{\begin{eqnarray}}
\def\ea{\end{eqnarray}}
\def\fun#1#2{\lower3.6pt\vbox{\baselineskip0pt\lineskip.9pt
        \ialign{$\mathsurround=0pt#1\hfill##\hfil$\crcr#2\crcr\sim\crcr}}}

\def\zp#1#2#3{{\it {Z. Phys.}} {\bf{C#1,}} #2 (#3)}

\def\prl#1#2#3{Phys. Rev. Lett. {\bf #1}, #2 (#3)}
\def\pr#1#2#3{Phys. Rev. D {\bf #1}, #2 (#3)}
\def\pl#1#2#3{Phys. Lett. {\bf #1B}, #2 (#3)}
\def\np#1#2#3{Nucl. Phys. {\bf B#1}, #2 (#3)}

\newcommand{\mpl}[2]{{ Mod. Phys. Lett.}     {\bf A#1}, #2 }

\newcommand{\hs}{\hat{s}^2}
\newcommand{\hc}{\hat{c}^2}

\begin{document}
\draft
\twocolumn[\hsize\textwidth\columnwidth\hsize\csname @twocolumnfalse\endcsname
\title{Z-Observables and the Effective Weak Mixing Angle in the MSSM}
\author{A. Dedes\cite{dedes}, K. Tamvakis}
\address{Division of Theoretical Physics, Physics Department,
University of Ioannina, GR-45 110, Greece}
\author{A. B. Lahanas}
\address{Physics Department, Nuclear and Particle Physics Section,
University of Athens, Athens 157 71, Greece}

\maketitle
\widetext

\begin{abstract}
\noindent
We make a comparison of the predicted effective weak mixing
angle, the Z-on resonance 
asymmetries and the W-boson mass to the 
LEP and SLD data at their present status.
 We find that the predicted MSSM values for
the effective weak mixing angle are in agreement with the LEP+SLD
average value for a ``heavy'' SUSY breaking scale while we observe
an agreement with SLD data in the case of a ``light'' SUSY breaking
scale. The resulting values for the W-boson mass and for the 
electron left-right asymmetries are compatible with CDF,UA2,D{\O} and
LEP data respectively. Unexpectedly we find that the
 supersymmetric QCD contributions to the Z-observables
tend to vanish everywhere in the $M_{1/2}$-$M_0$ plane. Furthermore,
values of $M_{1/2}$ which are greater than 500 GeV are favoured by the 
MSSM if one considers the current experimental
value for the strong coupling.
\end{abstract}

\vspace*{5mm}
]

\narrowtext

\section{Introduction}

The weak mixing angle is defined as a ratio of the gauge 
couplings
\begin{equation}
\hs(Q) = \frac{\hat{g}^{\prime 2}(Q)}{\hat{g}^2(Q)+\hat{g}^{\prime 2}(Q)}  \;,
\label{eq:1}
\end{equation}
where $\hat{g}$ and $\hat{g}^\prime$ are the $SU(2)$ and $U(1)_Y$
gauge couplings respectively. These couplings are running in the 
sense that they depend on the energy scale Q, and in supersymmetry
they are evaluated in the $\overline{DR}$ scheme. The weak mixing angle
$\hs$ can be easily predicted from a GUT model, {\it i.e.} SU(5) or
from a string derived model like the flipped SU(5). There are many
sources of the determination of the weak mixing angle in terms
of experimentally measurable quantities. From muon
decay for instance and knowing the most accurate parameters, 
$M_Z=91.1867 \pm 0.0020 GeV$, $\alpha_{EM}=1/137.036$ and 
$G_F=1.16639(1)\times10^{-5}\, GeV^{-2}$, we can define the $\hs$
at $M_Z$ by the expression
\begin{equation}
\hs \hc=\frac{\pi\, \alpha_{EM}\, 
}{\sqrt{2}\,M_Z^2\,G_F\,
(1-\Delta \hat{\alpha})\,\hat{\rho}\,(1-\Delta \hat{r}_W)}\; ,
\label{eq:2}
\end{equation}
where 
\begin{eqnarray}
\hat{\rho}^{-1}&=&1-\Delta \hat{\rho}=
1-\frac{\Pi_{ZZ}(M_Z^2)}{M_Z^2}+\frac{\Pi_{WW}(M_W^2)}
{M_W^2}\label{eq:3}\;,\\[3mm]
\Delta \hat {r}_W&=&\frac{\Pi_{WW}(0)-\Pi_{WW}(M_W^2)}{M_W^2}+
\hat{\delta}_{VB}\label{eq:4}\;,\\[3mm]
\hat{\alpha}&=&\frac{\alpha_{EM}}{1-\Delta{\hat{\alpha}}}
\label{eq:5}\;.
\end{eqnarray}
$\Pi$'s are the loop contributions to the transverse Z and W
gauge bosons self energies and they contain both logarithmic and 
finite parts. The quantity $\hat{\delta}_{VB}$ contains vertex+box
corrections and the term $\Delta \hat{\alpha}$  contributions from
light quarks and leptons as well contributions from heavy particle
thresholds {\it i.e.} top quark, Higgs bosons, superpartner masses.
The contributions (\ref{eq:3}),(\ref{eq:4}),(\ref{eq:5})
 include logarithmic corrections
of the form
\begin{eqnarray}
\log \biggl ( \frac{M_{1/2}^2}{M_Z^2} \biggr ) \;,
\nonumber
\end{eqnarray}
and if the soft gaugino mass $M_{1/2}$ takes on
large values we  refer to this quantity as  a ``large log" quantity.
We have proven in \cite{Dedes5} that the weak mixing angle behaves like
\begin{equation}
\hs \sim \log \biggl ( \frac{M_{1/2}^2}{M_Z^2} \biggr ) \;.
\label{eq:6}
\end{equation}
This is clear from Figure 1, where if we increase the value of
$M_{1/2}$ then $\hs$ takes on large values which means that large
logs have not been decoupled from its value. Through this note we
always assume the unification of gauge couplings up to two loops
  and the constraint
from the radiative symmetry breaking \cite{Tamvakis}
 where the full one loop minimization
conditions have been taken into account.
\begin{figure}
\centerline{\psfig{figure=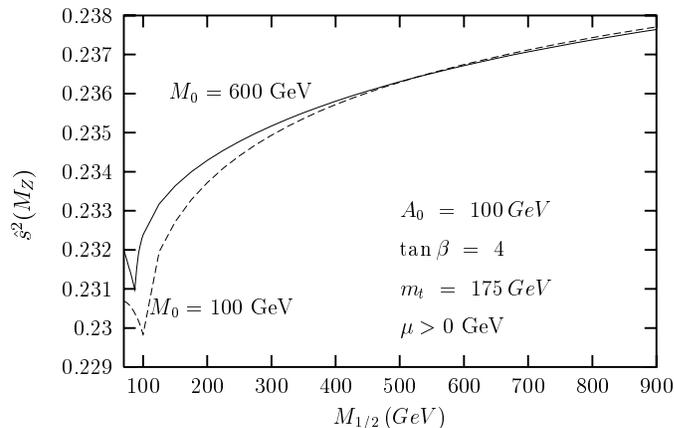,height=2.2in}}
\caption{The extracted value of $\hs$ as a function of $M_{1/2}$.
The logarithmic behavior resulting from the soft gaugino masses is obvious.}
\end{figure}
\twocolumn[\hsize\textwidth\columnwidth\hsize\csname @twocolumnfalse\endcsname

\widetext
\begin{table}
\caption{Partial and total contributions to ${\Delta k}_f$, 
($f=lepton,charm,bottom$), for two sets of inputs shown at the top.
Also shown are the predictions for the effective weak mixing angles and the
asymmetries. In the first five rows we display the universal contributions
to $10^3 \; \times \;{\Delta k}$ of squarks ($\tilde q$),
sleptons($\tilde l$), Neutralinos and Charginos (${\tilde Z},{\tilde C}$),
ordinary fermions and Higgses 
(The number shown in the middle below the
"charm" column refers to "lepton" and "bottom" as well). In the next five rows
we display the contributions of gauge bosons as well as the supersymmetric
$Electroweak \; (EW)$ and $SQCD$ vertex
and external fermion wave function renormalization corrections
to $10^3 \; \times \;{\Delta k}$. }
\begin{tabular}{ccccccccc}
 &$M_0=100$&$M_{1/2}=100$&$A_0=100$& &
 &$M_0=600$&$M_{1/2}=600$&$A_0=600$ \\
 &$m_t=175$&$tanb=4$&$\mu > 0$&  &
 &$m_t=175$&$tanb=4$&$\mu > 0$  \\
\tableline
\\
 &$lepton$ &$charm$ &$bottom$& &
 &$lepton$ &$charm$ &$bottom$ \\   \\
\tableline
\\
${\tilde q}$& &-4.3481 & && & &-10.0677 & \\
${\tilde l}$& &-0.2166 & && & &-0.4042 & \\
${\tilde Z},{\tilde C}$& &7.087 & && & &-12.2366 & \\
$Fermions$& &4.8177 & && & &4.4290 & \\
$Higgs$& &-0.4390 & && & &-1.1523 & \\  \\
$Gauge$&-3.2491  &-3.7027 &2.2999  &&
       &-3.0766   &-3.5184 &2.2771  \\ \\
$Vertex(EW)$&-1.9399&2.1535 &6.9837  &&
       &2.8229   &5.6099 &21.2526  \\
$Wave(EW)$&-0.1487&-2.210 &-8.3524 &&
       &-2.8280   &-5.6037 &-21.0202\\   \\
$Vertex(SQCD)$& - &0.1217 &-1.0491  &&
       & -  &0.2416 &-1.0202 \\
$Wave(SQCD)$&- &-0.1204 &1.0057 &&
       & -  &-0.2416 &1.0190\\
\tableline
\\ 
${\Delta k}\;(\times \; 10^2 ) $ &0.1557 &0.3137 &0.7782 &&
&-2.2513 &-2.2944 &-1.6424  \\
${sin^2} {\theta}_f$& 0.23019 & 0.23055 &0.23162 && &0.23145
& 0.23135 & 0.23277\\
${\cal A}_{LR}^f$& 0.1575 & 0.6708 &0.9355 && &0.1476 &0.6681 & 0.9347\\
${\cal A}_{FB}^f$& 0.0186 & 0.0792 &0.111 && &0.0163 & 0.0740 & 0.1035\\

\end{tabular}
\end{table}
]
\narrowtext

Although the weak mixing angle $\hs$ provides a convenient means to
test unification in unified extensions of the Standard Model (SM)
or the MSSM it is {\it not} an experimental quantity. 
There are several studies in the literature evaluating the weak mixing angle
both in the SM \cite{Sirlin,gambino,all1,all2,djoua} 
and in the MSSM \cite{polon}.

\section{The Effective Weak Mixing Angle}

LEP collaborations employ the effective weak mixing angle
$s_f^2$ first introduced by Degrassi and Sirlin \cite{Sirlin}.
Electroweak corrections to the $Zf\bar{f}$ vertex yield the
effective Lagrangian which defines the effective weak mixing angle as
\begin{equation}
s_f^2 \equiv \hs \hat{k}_f = \hs ( 1 + \Delta \hat{k}_f )\;,
\label{eq:7}
\end{equation}
or in analogous way as we define the eq.(\ref{eq:2})
\begin{equation}
s^2_{f}c^2_{f}=\frac{\pi\, \alpha_{EM}\, (1+\Delta \hat{k}_f)\,
(1-\frac {\hs}{\hc}\,\Delta  \hat{k}_f)}{\sqrt{2}\,M_Z^2\,G_F\,
(1-\Delta \hat{\alpha})\,\hat{\rho}\,(1-\Delta \hat{r}_W)}\;,
\label{eq:8}
\end{equation}
where

\begin{eqnarray}
\Delta \hat{k}_f&=&\frac{\hat{c}}{\hat{s}}\,\frac{\Pi_{Z\gamma}(M_Z^2)-
\Pi_{Z\gamma}(0)}{M_Z^2}+ 
\delta k_f^{SUSY}+\cdots \;.
\label{eq:9}
\end{eqnarray}
$\Pi_{Z\gamma}$ denotes the Z-$\gamma$ propagator corrections and 
the term $\delta k_f^{SUSY}$ the non-universal corrections resulting 
from the vertex and wave function renormalization corrections. 
The non-universal corrections $\delta k_f^{SUSY}$ contain both diagrams with
electroweak and supersymmetric QCD corrections. There are also other
SM contributions which although included in our numerical analysis,
 they are
irrelevant to the discussion here.

The effective weak mixing angle is an experimental measured 
quantity and its average LEP+SLD value is 0.23152 $\pm$ 0.00023 
\cite{Altarelli}. The difference between the two angles in the SM
is
\begin{eqnarray}
{\rm SM :}\;\; s_l^2-\hs \lesssim O(10^{-4})\;,
\nonumber
\end{eqnarray}
which is less than the error quoted by the experimental groups.
This fact is due to the small total contribution from fermions and
bosons occurring at the one loop level, in the 
$\overline{MS}$ scheme and the only dominant contributions are
obtained from the two-loop heavy top contributions and three-loop
QCD effects.

In the MSSM this picture is changed dramatically. The difference
between the two angles can be at one loop,
\begin{eqnarray}
{\rm MSSM :}\;\; s_l^2-\hs \lesssim O(10^{-3})\;.
\nonumber
\end{eqnarray}
So, in the case of the MSSM the two angles may have completely 
different values and the difference between them can be much
greater than the experimental error.
As we have shown in ref.\cite{Dedes5} although $\hs$ suffers from
large logs in the MSSM, this is not the case for the $s^2_f$.

So far the situation is clear : GUT analysis is able to predict
the weak mixing angle $\hs$ and then eq.(\ref{eq:7}) can translate the
result into the experimentally measured quantity, the effective weak
mixing angle $s_f^2$.

As we have previously mentioned, among the interesting features 
of the effective weak mixing angle is that large logs get decoupled
from its value. In ref.\cite{Dedes5} we have shown this fact both
analytically and numerically. Here we present only the results of
the analytical proof. Math formulae can be found in ref.\cite{Dedes5}.

\subsection{Analytical Proof of the Decoupling}

There are three sources of large logs which affect the value of the
weak mixing angle $s_f^2$:\\
1. Gauge boson self energies which feed large logs to the quantities
$\Delta \hat r_W$, $\hat\rho$ and $\Delta \hat k_f$. These corrections
are canceled against large logs stemming from the electromagnetic coupling
$\hat{\alpha}(M_Z)$.\\
2. Vertex, external wave function renormalizations and box corrections
to muon decay which affect $\Delta \hat r_W$ through $\delta_{VB}^{SUSY}$.
These corrections are cancelled against each other out.\\
3. Non-universal vertex and external fermion corrections to 
$Z \overline{f} f$ coupling which affects $\Delta \hat k_f$. 
These corrections are also cancelled against each other out.\\
We conclude that $s_f^2$ behaves like $\sim O(\frac{M_Z}{M_{1/2}})$ which
is the SUSY contribution from the finite parts. It seems that SUSY follows
the well known Appelquist-Carazzone decoupling theorem \cite{Carazzone}.

\subsection{Numerical Proof of the Decoupling}

In order to prove the decoupling of large logs, we have included
the full supersymmetric one loop corrections for the universal and
non-universal part of eq.(\ref{eq:8}). 
In all figures that follows we assume the constraint 
from the radiative EW symmetry breaking \cite{Tamvakis}
 and the universality for soft
breaking masses at the unification scale $M_{GUT}$.
In Figure 2 we plot the leptonic effective weak mixing angle $s_l^2$ as
a function of the soft gaugino mass parameter $M_{1/2}$ for a low
($M_0=100$ GeV) and high ($M_0=600$ GeV) value of the soft scalar
parameter, $M_0$. 
\begin{figure}
\centerline{\psfig{figure=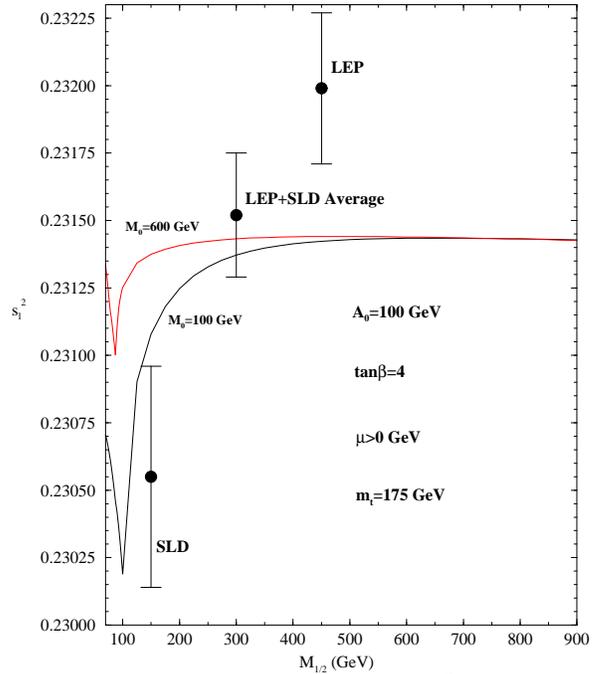,height=3.5in}}
\caption{The effective weak mixing angle $s_l^2$ versus $M_{1/2}$
for two characteristic values of $M_0$. The top quark mass is fixed 
at $m_t=175$ GeV.}
\end{figure}
We obtain that increasing the value of $M_{1/2}$, $s_l^2$ takes 
on constant values in the limit $M_{1/2}\rightarrow 900$ GeV independently
on the soft squark mass $M_0$. Finite corrections of the form 
$O(\frac{M_Z}{M_{1/2}})$ are significant in the region $M_{1/2}\sim M_0 
\lesssim 200$ GeV with the tendency to decrease (increase) the $s_l^2$ when
$M_{1/2}$ is greater(smaller) than $M_Z$. In addition lowering $M_0$ we
obtain smaller values for the $s_l^2$. It is worth noting that
MSSM seems to agree with the average LEP+SLD value while in the
vicinity of $M_0 \sim M_{1/2} \simeq 100$ GeV, $s_l^2$ agrees with the
SLD experimental value\footnote{The LEP average $s_{eff}^2=0.23199\pm 0.00028$
differs by 2.9$\sigma$ from SLD value $s_{eff}^2=0.23055\pm 0.00041$
obtained from the single measurement of left-right asymmetry.}.
The parameters $A_0$ and $\tan\beta$\footnote{The 
independent parameter $\tan\beta$ plays an important role
in the case of $s_b^2$.} do not affect significantly the
extracted value of $s_l^2$. 

Going deeper, we display in Table I the partial and total
contributions to $\Delta k_f$. Also shown are the predictions
for the effective weak mixing angle $s_f^2$ and the asymmetries.

We would like to stress five interesting points on this Table :

$\bullet$ The bulk of the SUSY corrections to $\Delta \hat{k}_f$ is
carried by the universal corrections which are sizeable due to their 
dependence on large log terms.

$\bullet$ The contribution of Higgses which is small mimics that of the
SM with a mass in the vicinity of $\sim 100$ GeV.

$\bullet$ Gauge\footnote{Gauge boson contribution are different for the
different fermion species $l,c,b$. This is due to the fact that their
non-universal corrections depend on the charge and weak isospin
assignments of the external fermions. In the case of the bottom
quark there are additional, significant top quark corrections.}
and Higgs boson contributions
tend to cancel large universal contributions of matter fermions.

$\bullet$ We find that the non-universal Electroweak SUSY
corrections are very small for a ``heavy'' 
SUSY breaking scale, $M_{1/2}=M_0=A_0=600$ GeV.
Although separetaly vertex and external fermion corrections are
large they cancel each other out yielding contributions almost two
orders of magnitude smaller than the rest of the EW corrections. For a
``light'' SUSY breaking scale, $M_{1/2}=M_0=A_0=100$ GeV this
cancellation does not occur due to the large splittings of the squarks 
and sleptons.

$\bullet$ The non-universal SQCD contributions, although a priori
expected to be larger than the electroweak corrections turn out to
be even smaller. This fact is due to cancellations both from its 
diagram in vertex or in wave function renormalization, and from 
different types of the diagrams. This situation is valid both
in the ``heavy'' and ``light'' SUSY breaking limit and we believe
that this cancellations arise from a QED like Ward identities between
vertex and wave function renormalization corrections.

\section{Z-Boson Observables in the MSSM}

Having obtained the value of the effective weak mixing angle, 
we are able to find other Z-observables in the MSSM. For 
instance the physical W-boson mass is defined by
\begin{equation}
M_W\ =\ M_Z~\hat{c}~\sqrt{\hat{\rho}} = 
M_Z~ c_f~ \sqrt{\frac{\hat{\rho}}{1-
\frac{\hs}{\hc} \Delta {\hat k}_f}}\;.
\label{eq:10}
\end{equation}

In figure 3 we plot the W-boson mass versus the input values
for the $M_{1/2}$ parameters for two characteristic values 
of $M_0$ in the ``light'' SUSY breaking scale ($M_0=100$ GeV)
and in the ``heavy'' SUSY breaking scale ($M_0=600$ GeV).
As one can see the W-mass is in agreement with the presently 
experimentally observed value, $M_W=80.427\pm 0.075$($M_W=80.405\pm 0.089$)
GeV obtained from LEP(CDF,UA2,D\O) experiments \cite{booklet} for
rather low(high) values of $M_{1/2}$. Moreover, our results are in
agreement with those of refs.\cite{Bagger,Hollik1,Pok3}.

\begin{figure}
\centerline{\psfig{figure=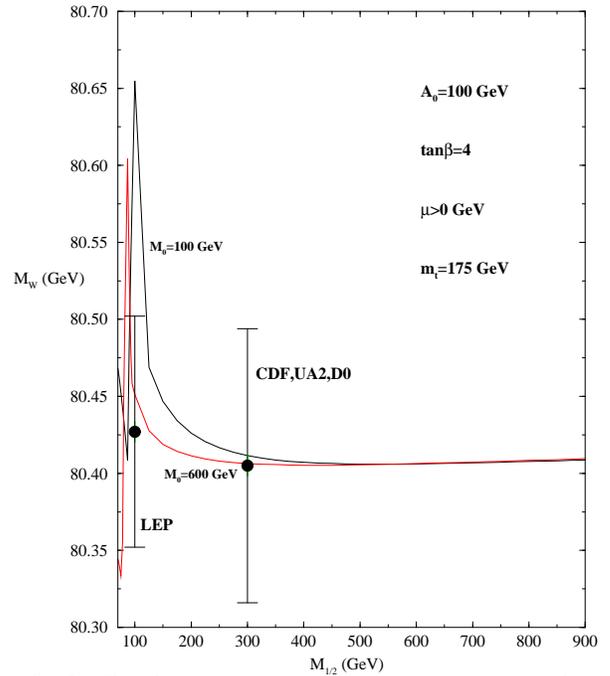,height=3.5in}}
\caption{Predicted values for the W-boson mass in the MSSM.}
\end{figure}

Having at hand the effective weak mixing angle we can
easily derive  the left-right asymmetries 
which in term of the $s_f^2$ can defined as
\begin{equation}
A_{LR}^f \ =\ {\cal A}^f \ =\ \frac{2~T_3^f~\left (T_3^f-2~s_f^2~Q^f
\right )}{T_3^{f 2} + \left (T_3^f - 2 s_f^2 Q^f \right )^2} \;,
\label{eq:11}
\end{equation}
where $Q^f$ is the electric charge and $T_3^f$ is the third component 
of the isospin of the fermion $f$. As it is depicted in Figure 4, the
MSSM predictions for ${\cal A}_e$ agree with LEP+SLD average value
(${\cal A}_e = 0.1505 \pm 0.0023$) when both $M_{1/2}$ and $M_0$
take on values around $M_Z$. In the heavy SUSY breaking scale, the
MSSM agrees with the LEP value, ${\cal A}_e = 0.1461 \pm 0.0033$. Note
that increasing the value of $M_{1/2}$, ${\cal A}_e$ takes on constant 
values which means that large logs have been decoupled from the
expression (\ref{eq:11}). In this case we have recovered  
SM results displayed in Particle Data Booklet \cite{booklet}.

\begin{figure}
\centerline{\psfig{figure=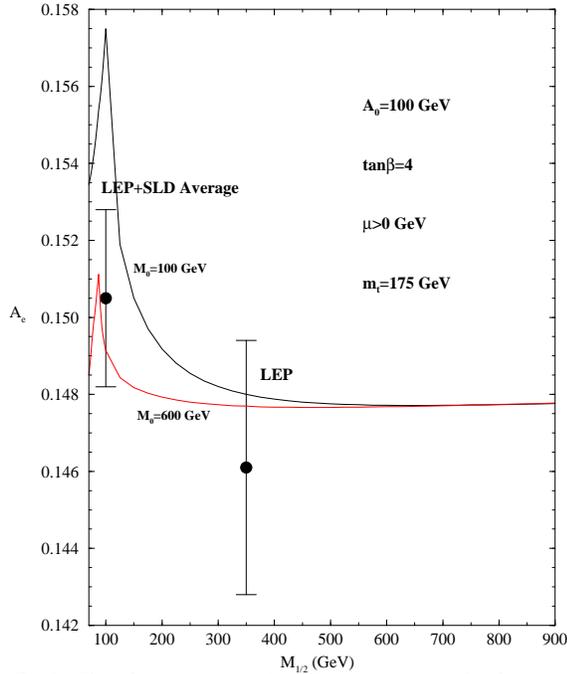,height=3.5in}}
\caption{Predicted values for the electron left right asymmetry
 in the MSSM.}
\end{figure}

So far we have not considered the constraint resulting from
the experimental value of $\alpha_s(M_Z)$ \cite{dedes2}.
In Figure 5 we have
plotted the acceptable values of the soft breaking parameters
which are compatible with LEP ($\alpha_s(M_Z)=0.119\pm 0.004$, 
$s_{eff}^2=0.23152 \pm 0.00023$)\cite{Altarelli}
 and the CDF/D\O ($m_t=175 \pm 5$ GeV) data
\cite{top}.

\begin{figure}
\centerline{\psfig{figure=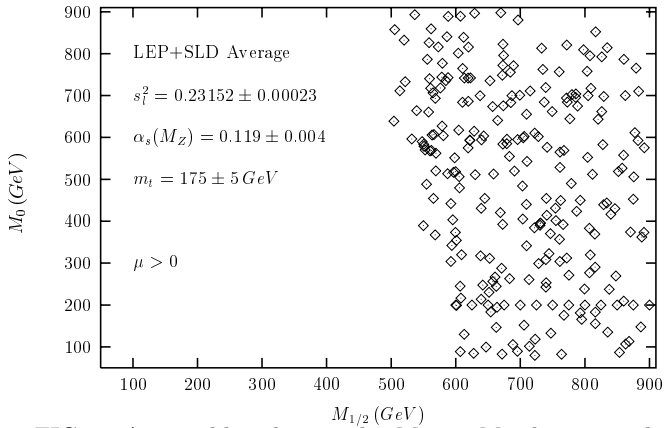,height=2.2in}}
\caption{Acceptable values in the $M_{1/2}-M_0$ plane according to the
LEP+SLD data. Only large values of $M_{1/2}$ are acceptable.}
\end{figure}

 The trilinear soft couplings as well as the 
parameter $\tan\beta (M_Z)$ are taken arbitrarily in the region
(0-900 GeV) and (2-30), respectively.
 As we observe from Figure 5,
MSSM with unification of gauge coupling constants, universality
of the soft masses at $M_{GUT}$ and radiative EW symmetry breaking
is valid in the region $M_{1/2}\gtrsim 500$ GeV and $M_0\gtrsim 70$ GeV.
The lower bound on $M_{1/2}$ is put mainly from the experimental value
of $\alpha_s$ and on $M_0$ from the requirement that LSP is neutral.
In this region, the physical gluino mass is above 1 TeV, the LSP is
$\gtrsim 200$ GeV, the squark and slepton masses are $\gtrsim 800$ GeV 
and $\gtrsim 210$ GeV and the light Higgs boson mass is greater than
$108$ GeV.

\section{Conclusions}

In ref.\cite{Dedes5} we have included the full one-loop 
supersymmetric corrections to the effective weak mixing angle
which is experimentally determined in LEP and SLD experiments.
Our analysis enables one to pass from the ``theoretical'' weak mixing
angle $\hs$, which can be predicted from GUT analysis, to
the experimental effective weak mixing angle $s_f^2$.
We conclude that :

$\bullet$ There are no logarithmic corrections of the form $\log 
(\frac{M_{1/2}^2}{M_Z^2})$ (or more generally $\log
(\frac{M_{SUSY}^2}{M_Z^2})$) to the effective weak mixing angle.

$\bullet$ Supersymmetric QCD corrections tend to vanish everywhere in
the $M_{1/2}-M_0$ plane.

$\bullet$ The predicted MSSM values for the effective weak mixing
angle are in agreement with the present LEP+SLD average value in the
``heavy'' SUSY breaking scale while they are in agreement with the
SLD data in the ``light'' SUSY breaking scale..

$\bullet$ MSSM predicts values of the W-boson mass which are in
agreement with the new CDF,UA2,D{\O} average value.

$\bullet$ In the ``heavy'' SUSY breaking limit MSSM seems to
prefer the experimental LEP value for the electron left-right asymmetry
${\cal A}_e$.

$\bullet$ Finally, values of $M_{1/2}$ which are greater than
$500$ GeV are favoured by the MSSM if one assumes the present LEP+SLD
and CDF/D{\O} data for $s_{eff}^2$, $\alpha_s$ and $m_t$, respectively.

\acknowledgments

 A.D. and K.T. acknowledge financial support from the 
research program $\Pi{\rm ENE}\Delta$-95
of the Greek Ministry of Science and Technology. A.B.L. and K. T. acknowledge
 support from
the TMR network ``Beyond the Standard Model", ERBFMRXCT-960090. 
A. B. L. acknowledges 
support from the Human Capital and Mobility program CHRX-CT93-0319.

\end{document}